\documentclass[12pt]{article}
\usepackage{amsfonts}
\usepackage{epsfig}

\newcommand{\eq}{\begin{equation}}
\newcommand{\eqx}{\end{equation}}
\newcommand{\eqn}{\begin{eqnarray}}
\newcommand{\eqnx}{\end{eqnarray}}
\newcommand{\f}[2]{\frac{#1}{#2}}

\newcommand{\tr}{\mbox{\rm tr}\,}

\newcommand{\Dl}{\Delta}
\newcommand{\lm}{\lambda}
\newcommand{\Lm}{\Lambda}
\newcommand{\om}{\omega}
\newcommand{\al}{\alpha}

\renewcommand{\th}{\theta}
\newcommand{\qqqq}{\quad\quad\quad\quad}
\newcommand{\qqq}{\quad\quad}

\newcommand{\ZZ}{{\mathbb Z}}

\newcommand{\NN}{{\cal N}}

\newcommand{\cor}[1]{\left\langle{#1}\right\rangle}

\title{Phases of $\NN=1$ theories and factorization of Seiberg-Witten
  curves\footnote{Talk given at the XLV School of Theoretical Physics,
  Zakopane, 3-12 June 2005.}}

\author{Romuald A. Janik\thanks{e-mail:{\tt ufrjanik@if.uj.edu.pl}} \\ \\
Institute of Physics\\
Jagellonian University,\\
ul. Reymonta 4, \\
30-059 Krak{\'o}w\\
Poland
}
\begin{document}

\maketitle

\begin{abstract}
In this talk I review the structure of vacua of $\NN=2$ theories
broken down to $\NN=1$ and it's link with factorization of
Seiberg-Witten curves. After an introduction to the structure of vacua
in various supersymmetric gauge theories, I discuss the use of the
exact factorization solution to identify different dual
descriptions of the same physics and to count the number of connected
domains in the space of $\NN=1$ vacua.
\end{abstract}

\section{Introduction}

Supersymmetric gauge theories serve as a laboratory for studying
nonperturbative physics of gauge theories. Thanks to the huge
simplifications and constraints on the possible behaviour of these
theories due to the additional (super-)symmetry present, it is
possible to obtain exact results and obtain fully nonperturbative
answers to at least some questions. 

Remarkable progress has been made in exploring the structure of vacua
of various supersymmetric gauge theories. Since the early progress on
$\NN=1$ SYM theories in the eighties (see e.g. the review talk
\cite{SHIFMAN} and references therein), 
further elaborated by emphasizing the
constraints of holomorphicity in the nineties \cite{SEIBHOL}  a very
complete picture 
was obtained in the case of $\NN=2$ theories \cite{SW1,SW2} 
bringing together such
mathematical structures as algebraic curves and integrable
systems. More recently string theory constructions led to progress in
studying $\NN=2$ theories broken down to $\NN=1$ through an addition
of a superpotential for the adjoint superfield in the original $\NN=2$
theory \cite{V2}, and a remarkable link with random matrix theory \cite{DVP}. 
Later this link was understood purely in field theoretic terms
\cite{FERRARI1,DVZANON,CDSW}. 

Later \cite{FERRARI2,PHASES1} it was found
that the structure of the $\NN=1$ vacua where the gauge group is
broken down to $U(N_1) \times U(N_2)$ is surprisingly complicated
and they lie in connected domains, whose number jumps widely from
one value of $N_c=N_1+N_2$ to another.

In this talk I would like to give an introduction to these issues and
show how one can study the structure of these vacua a.k.a. `the phases of
$\NN=1$ theories' in a systematic fashion using an exact general
solution for the factorization of Seiberg-Witten curves in the above
situation \cite{FACT}. 

The plan of this talk is as follows. First I would like to present the
structure of vacua for a range of SYM starting from pure $\NN=1$
through $\NN=2$ to $\NN=2$ broken down to $\NN=1$. Then I discuss the
interrelation between the $\NN=1$ vacua and factorization of
Seiberg-Witten curves, in section 4 I review the solution of the
factorization problem given in \cite{FACT} and proceed to apply it to
determine connected domains of $\NN=1$ vacua. I close with a summary
and outlook. 

\section{The vacua of supersymmetric gauge theories}

In this section I will briefly review the structure of the vacua of
some supersymmetric gauge theories starting from the simplest case and
ending on the theories which are the focus of this talk.

\subsection*{Pure $\NN=1$ Supersymmetric Yang-Mills (SYM)}

Pure $\NN=1$ SYM is the simplest supersymmetric gauge theory. Apart
from the ordinary Yang-Mills field, it contains just a minimally
coupled fermion in the {\em adjoint} representation. This fermion
is called `the gaugino' or `gluino'. The SYM lagrangian is just
\eq
\label{e.lsym}
L=-\f{1}{4g^2} F^a_{\mu\nu}F^{a\mu\nu} +\f{1}{g^2} \bar{\lm}^a i
D\lm^a+ i\f{\th}{32\pi^2} F^a_{\mu\nu} \tilde{F}^{a\mu\nu}
\eqx 
where we included also the topological $\th$-term. 

Let us now explore
the symmetries of this theory. On the classical level (\ref{e.lsym})
is invariant under a $U(1)$ $R$-symmetry which only transforms the gaugino
\eq
\lm^a \to e^{i\al} \lm^a
\eqx
This symmetry is broken on the quantum level by an anomaly. However it
may be compensated by a shift in the $\th$ angle: $\th \to \th+2N_c
\al$. Then since a shift of $\th$ by $2\pi$ gives the same theory one
is left with a residual discrete $\ZZ_{2N_c}$ symmetry.

It turns out that in this theory the gaugino condenses
$\cor{\lm\lm}=\Lm^3$. This vacuum expectation value (VEV) breaks down
$\ZZ_{2N_c}$ down to $\ZZ_2$. The remaining $N_c$ transformations move
between $N_c$ vacua characterized by
\eq
\cor{\lm\lm}=\Lm^3 e^{i\f{k}{N_c}}  \qqqq k=0 \ldots N_c-1
\eqx
So a $U(N_c)$ SYM has $N_c$ discrete vacua labelled by $k=0\ldots
N_c-1$. Around each such vacuum the $SU(N_c)$ part of the theory will
develop a mass gap, and one will be left with a massless $U(1)$.

Analogous reasoning for a SYM theory with gauge group $U(N_1)
\times U(N_2)$ leads to $N_1 \cdot N_2$ vacua labelled by two integers
$k_1=0\ldots N_1-1$ and $k_2=0\ldots N_2-1$. The massless degrees of
freedom will be $U(1)^2$.

\subsection*{$\NN=2$ SYM and Seiberg-Witten curves}

The $\NN=2$ SYM is just the $\NN=1$ SYM together with an adjoint
chiral superfield (containing an adjoint scalar $\phi$, and fermionic
partner). The vacuum condition is just that $\phi$ can be
diagonalized, so one has an $N_c$ dimensional moduli space of
vacua. These can be parametrized by $N_c$ complex parameters
\eq
u_p\equiv \cor{\f{1}{p} \tr \phi^p} = \f{1}{p} \sum_{i=1}^{N_c} x_i^p
\qqqq p=1 \ldots N_c 
\eqx 
Seiberg and Witten found \cite{SW1,SW2} that all the low energy
properties of the theory are encoded in the geometry of a certain
Riemann surface called the Seiberg-Witten curve. I will now review
this some features of this construction.

Around a generic vacuum (a generic point in the moduli space) one will
have $U(1)^{N_c}$ low energy effective theory. Its couplings are
encoded in the geometry of the associated Seiberg-Witten curve
\eq
\label{e.sw}
y^2=P^2_{N_c}(x;\{u_p\})-4\Lm^{2N_c}
\eqx
where $\Lm$ is the scale of the theory while $P_{N_c}(x;\{u_p\})$ is
a polynomial of order $N_c$ given by
\eq
P^2_{N_c}(x;\{u_p\})=(x-x_1)(x-x_2)\ldots (x-x_{N_c})
\eqx
A lot of other physical properties of the theory are encoded in the
Seiberg-Witten curve (\ref{e.sw}). In particular there are $N_c-1$
species of monopoles which are generically massive. They become
massless at special points in the moduli space of vacua where the
$u_p$'s are tuned so that the right hand side of the Seiberg-Witten
curve (\ref{e.sw}) has double zeroes. We say then that the Seiberg-Witten
curve {\em factorizes}. I will consider below the case when all but one
species of monopoles become massless. Then the SW curve can be written
as
\eq
\label{e.swfact}
y^2=P^2-4\Lm^{2N_c}=F_4(x) \cdot H_{N_c-2}^2(x)
\eqx
where $F_4(x)$ is a polynomial of degree 4. The factorization problem
which we will consider below is to find the $\{ u_p \}$'s for which
the SW curve looks like (\ref{e.swfact}).

A final piece of information that one can extract from the
Seiberg-Witten curve (\ref{e.sw}) is the knowledge of all VEV's for
$\tr \phi^k$. This is in fact nontrivial as for $k>N_c$ one has
instanton corrections to the classical VEV's (equal to $\sum_i
x_i^k$). The answer is
conveniently expressed in terms of the meromorphic 1-form
\eq
\om=\f{d}{dx} \log(P+\sqrt{P^2-4\Lm^{2N_c}})\,
dx=\f{P'}{\sqrt{P^2-4\Lm^{2N_c}}}\, dx
\eqx 
Then the VEV's are given by
\eq
\cor{\tr \phi^k} =res_{x=\infty}\, x^k \cdot \om
\eqx

\subsection*{$\NN=2$ broken down to $\NN=1$}

The final theory whose vacuum structure I would like to describe is
the $\NN=2$ theory which is broken down to $\NN=1$ through the
addition of a superpotential term to the action:
\eq
\int d^4x \int d^2\th\, W_{tree}(\Phi)
\eqx
where $W_{tree}$ is a polynomial
\eq
W_{tree}(\Phi)=\sum_p g_p \f{1}{p} \tr \Phi^p
\eqx
This term breaks $\NN=2$ to $\NN=1$ and moreover modifies the vacuum
structure of the theory. Here we will just look at the classical
picture leaving the discussion of the full quantum description to the
next section.

The addition of the superpotential gives a condition for the classical
vacua 
\eq
W'_{tree}(\phi)=0
\eqx 
This means that the eigenvalues of $\phi$ should be distributed among
the extrema of $W_{tree}$. If all eigenvalues sit at a single extremum,
$U(N_c)$ is unbroken by the VEV of $\phi$ and at low energies we
should have a $U(N_c)$ pure $\NN=1$ SYM theory which leads to $N_c$
discrete vacua and leaving a massless $U(1)$. 

If $N_1$ eigenvales sit at one
extremum and the remaining $N_2$ at another, the gauge group is broken
down to a product $U(N_1)\times U(N_2)$ and one is left with $N_1
\cdot N_2$ vacua labeled by $k_1,k_2$. The massless degrees of freedom
left after gaugino condensation will be a $U(1)^2$ gauge theory.

This classical picture will have to modified on
the quantum level. In particular, as discussed above, one cannot treat
the eigenvalues of $\phi$ as strictly classical quantities (e.g. the
VEV's of higher powers of $\phi$ get instanton corrections). In the
next section we will reanalyze what happens once we take into account
our knowledge of the undeformed $\NN=2$ theory encoded in the
Seiberg-Witten curve.

\section{$\NN=1$ vacua and Seiberg-Witten curves}

The new ingredient which one has to include in order to describe the
vacua of $\NN=2$ broken down to $\NN=1$ are monopoles. If we neglect
them then the F-flatness conditions for the vacuum cannot be
met. E.g. if we just add a mass term for $\phi$:
\eq
W_{tree}=\f{1}{2} m\, \tr \phi^2=m \cdot u_2
\eqx
then varying w.r.t. $u_2$ gives $m=0$! The only way to obtain a
solution is to realize that in the superpotential one has to include
{\em all} relevant low energy degrees of freedom - in particular we
have to include monopoles which can become massless, and thus have to
be necessarily included in the low energy effective theory. So one has
to consider 
instead\footnote{Here we just quote for simplicity the $SU(2)$ case.}
\eq
W=m_{monopole}(u_2) M\tilde{M}+m \cdot u_2
\eqx 
Varying that superpotential we find that i) the value of $u_2$
(i.e. the point in the moduli space of the original $\NN=2$ theory)
has to be such that the monopole is massless and ii) the monopoles
condense. Recall from the previous section that the condition (i)
means that the Seiberg-Witten curve has to {\em factorize}.

The above is true in general. The Seiberg-Witten curve corresponding
to $\NN=2$ theory broken down to $\NN=1$ by the addition of a tree
level superpotential has to factorize. In case of unbroken gauge
group, all monopoles are massless, while for $U(N_c) \to U(N_1)\times
U(N_2)$, all but one are massless and the SW curve has to look like
(\ref{e.swfact}). The nontrivial branch cuts of the curve intuitively
correspond to the fact that at nonzero $\Lm$, the eigenvalues no
longer behave like classical quantities sitting exactly at the minima
of the superpotential.

\begin{figure}[t]

\centerline{\psfig{file=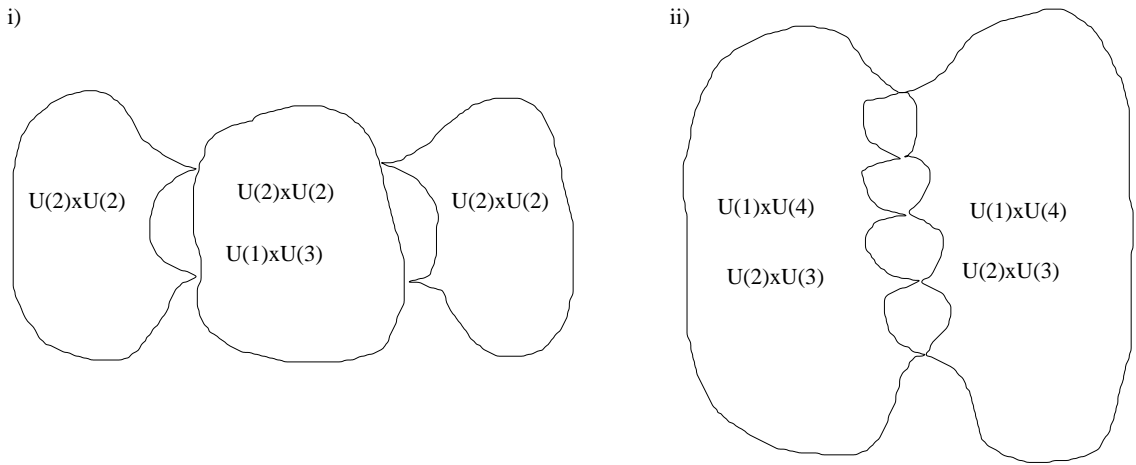,width=15cm}}

\caption{Connected domains of $\NN=1$ vacua for i) $N_c=4$, ii)
  $N_c=5$. The labels denote different dual descriptions within each
  domain.} 
\end{figure}   

The above observation, together with the fact that at low energies the
massless degrees of freedom are just $U(1)^2$ led to a surprising
possibility \cite{PHASES1} that at nonzero $\Lm$ one could not really distinguish
between certain breakings $U(N_c) \to U(N_1) \times U(N_2)$ (in some
specific discrete vacuum labelled by $k_1$ and $k_2$) and
$U(N_c) \to U(N'_1)  \times U(N'_2)$ in the vacuum labelled by $k'_1$
and $k'_2$. These could be interpreted as dual descriptions of the
same physics i.e. of the same underlying Seiberg-Witten curve.
Moreover it was found \cite{PHASES1} that once one changes the
deforming tree-level superpotential the vacua move around but {\em
  remain} within a certain number of connected domains (see fig. 1).
In \cite{PHASES1} 
an analysis was performed of the structure of these domains up to
$N_c=6$ by direct factorization of the Seiberg-Witten curves. In this
talk I would like to review a general solution of the factorization
problem valid for any $N_c$ and the resulting picture of the domains
of $\NN=1$ vacua. But before I present further details I will review
how are the parameters of the vacua $N_1$, $k_1$, $k_2$ linked to the
factorized SW curve.

In the case of (\ref{e.swfact}), there are two nontrivial compact
cycles $A$ and $B$. The periods of the meromorphic 1-form $\om$ are
then exactly $N_1$ and $\Dl k\equiv k_2-k_1$ (\cite{PHASES1,PHASES2}
see also \cite{FERRARI2}):
\eq
\f{1}{2\pi i} \oint_A \om =N_1 \qqqq \f{1}{2\pi i} \oint_B \om =\Dl k
\eqx
The fact that one can have dual descriptions is easy to understand
once one takes into account that the {\em choice} of $A$ and $B$
cycles is not unique and can be changed by a modular
transformation. Therefore $N_1$ and $\Dl k$ can change while the
underlying physics remains the same. However in order to have a
complete picture of the possible dual descriptions one has to
incorporate the remaining discrete parameter $k$ and look what sets of
discrete parameters $(N_1,\Dl k,k)$ lead to the same Seiberg-Witten
curve. These sets will just be the different possible dual
descriptions of the same physics. Each such set, on the other hand,
corresponds to a connected
domain of vacua\footnote{This is so, since in addition to the discrete
  parameters we will have continous parameters, varying which will
  fill out the domains.} as in fig. 1.

In the next section we will briefly present the solution to the
factorization problem.

\section{The solution to the factorization problem}

The factorization problem is to find the set of $\{u_p\}$'s in the
$\NN=2$ moduli space where the Seiberg-Witten curve factorizes as in
(\ref{e.swfact}). The case of complete factorization corresponding to
unbroken gauge group and {\em all} monopoles being massless was solved
by Douglas and Shenker \cite{DS} using special properties of Chebyshev
polynomials. That case would correspond to the $F_4$ in
(\ref{e.swfact}) being substituted by a degree two polynomial
$F_2$. Their solution, for each $N_c$ depends on a discrete parameter
$k=0,\ldots N_c-1$ which just labels pure $\NN=1$ vacua.

In the present case the situation is much more complicated. For each
$N_c$ the set of allowed labels (discrete parameters) increases (these
are $(N_1,k_1,k_2)$) and new types of vacua tend to appear which
cannot be related to those at smaller $N_c$ (see \cite{PHASES1} -- these
are the `Coulomb vacua' in the terminology of that paper).

The key to finding a solution to this problem are certain properties
of the meromorphic 1-form $\om$:

\begin{itemize}
\item For a factorized Seiberg-Witten curve of the form
  (\ref{e.swfact}), $\om$ defines a meromorphic 1-form on the {\em
  elliptic} curve
\eq
y^2=F_4(x)
\eqx
\item $\om$ has residues $\pm N_c$ at infinity
\item $\om$ has integer periods given by $N_1$ and $\Dl k$
\end{itemize}

Once we know a meromorphic 1-form on the elliptic curve $y^2=F_4(x)$
satsfying the above properties, we may reproduce the {\em full}
Seiberg-Witten curve i.e. find all the $\{ u_p\}$'s and the scale of
the theory $\Lm$ from further properties of $\om$:
\begin{itemize}
\item The $\{ u_p \}$ is given by
\eq
\label{e.upom}
u_p=\f{1}{p} \cdot res_{x=\infty} x^p \om
\eqx
\item The scale of the theory can be obtained from
\eq
\label{e.lmom}
\left\{\int_a^\infty \om \right\}_{reg}  \equiv \lim_{x\to\infty}
\left(\int_a^x \om - N_c \log x \right)=-\log \Lm^{N_c}
\eqx
where $a$ is a branch point of the Seiberg-Witten curve.
\end{itemize}

\begin{figure}[t]

\centerline{\psfig{file=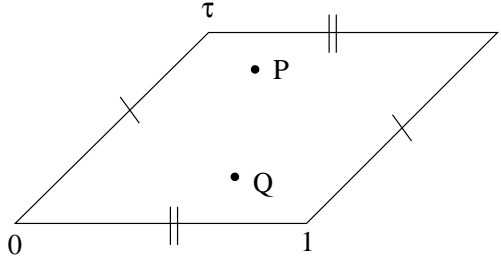}}

\caption{The torus (elliptic curve) represented as a pararellogram
  with identified opposite edges. $\tau$ is the modular parameter, $P$
  and $Q$ represent the infinities in the $(x,y)$ picture. They are
  related to the $a_i$'s appearing in (\ref{e.omtheta}) by
  $a_i+\f{1+\tau}{2}$.} 
\end{figure}

The key difficulty in finding a solution is an explicit construction
of a meromorphic 1-form 
with integer periods on an elliptic curve. In the conventional
representation $y^2=F_4(x)$ this seems impossible, however it is very
simple when one uses the representation of an elliptic curve as a
torus i.e. a pararellogram with identified edges (see fig. 2). Furthermore
one has to pick two points to represent the two infinities in the
$(y,x)$ representation. Then a meromorphic 1-form with prescribed poles at
these points with residues $\pm N_c$ is \cite{MUMFORD}
\eq
\label{e.omtheta}
\om=N_c \f{d}{dz} \log \f{\th(z-a_1)}{\th(z-a_2)} +C
\eqx  
where $\th(z)$ is the Jacobi theta function.
Without loss of generality we may set $a_1=0$. Then $a_2$ and $C$ are
uniquely determined from the periods, which in this representation of
the elliptic curve are trivial to calculate using quasiperiodicity of
the theta functions
\eqn
\f{1}{2\pi i} \int_A \om &=& \f{1}{2\pi i} \int_0^1 \om  = \f{C}{2\pi i} \\
\f{1}{2\pi i} \int_B \om &=& \f{1}{2\pi i} \int_0^\tau \om  = N_c
(a_1-a_2)+\f{C\tau}{2\pi i} 
\eqnx
and we easily get
\eqn
C &=& 2 \pi i N_1\\
a_2 &=& \f{N_1 \tau -\Dl k}{N_c}
\eqnx
Now we are almost done. In order to reproduce the $\{u_p\}$'s and
$\Lm$ we have to know how the original $x$ coordinate depends on
$z$. This can be done (see \cite{FACT} for details) and the result is
\eq
x(z)=\f{d}{dz} \log \f{\th(z-a_1)}{\th(z-a_2)}
\eqx
Using this formula one can readily calculate the $\{u_p\}$'s and
$\Lm$ from (\ref{e.upom})-(\ref{e.lmom}).
In fact $x(z)$ is unique only up to a linear transformation $x\to \al
x+x_0$. Using $\al$ we may set $\Lm^{2N_c}$ to its {\em given}
physical value. We are therefore left with a continous free parameter
$x_0$ and a {\em discrete} one coming from the fact that we are still
free to perform rescalings $x\to \al x$ with
\eq
\al=e^{2\pi i \f{k}{2N_c}} \qqqq k=0\ldots 2N_c-1
\eqx 
which do not modify $\Lm^{2N_c}$.

Putting all of this together we see that our solution of the
factorization problem is labelled by two complex continous parameters $\tau$
and $x_0$ and three {\em discrete} ones: $(N_1,\Dl k, k)$ i.e. exactly
the needed number to describe the expected $\NN=1$ vacua.

\section{Connected domains of $\NN=1$ vacua}

We can now ask the key questions about the structure of $\NN=1$ vacua:
\begin{itemize}
\item What are the possible dual descriptions of the same physics?
\item How many connected domains of vacua do we have for given $N_c$?
\end{itemize}

As described at the end of section 3, once we have an explicit
construction of the factorized Seiberg-Witten curve labelled by the
discrete parameters $(N_1,\Dl k,k)$, the above questions may be
translated into the questions: i) what {\em distinct} parameters
$(N_1,\Dl k,k)$ lead to {\em identical} Seiberg-Witten curves? ii)
once we identify the dual descriptions, how many distinct
Seiberg-Witten curves we are left with.

Some analysis of the properties of the theta functions under modular
transformations leads to the following identifications:
\eqn
\label{e.tri}
(N_1,\Dl k,k)\!\! &\equiv &\!\! (N_1,\Dl k-N_1,k) \qqq N_1 \leq \Dl k   \\
(N_1,\Dl k,k)\!\! &\equiv &\!\! (N_1,\Dl k-N_1+N_c,k+N_1+N_c \,\mbox{\rm mod }
2N_c) \; N_1 > \Dl k   \\
\label{e.trii}
(N_1,\Dl k,k)\!\! &\equiv &\!\! (N_c-\Dl k,N_1,k-N_1+N_c\, \mbox{\rm
  mod } 2N_c) \qqq \Dl k \neq 0
\eqnx
and their inverses. Of course the transformations of just $N_1$ and
$\Dl k$ under modular transformations are known from the outset,
however we need here to know exactly how does the third discrete
parameter $k$ enter these transformations.

\begin{table}[t]
\centerline{\begin{tabular}{|r|r||r|r|}
\hline
$N_c$ & No. of domains & $N_c$ & No. of domains \\
\hline
3 & 2 & 13 & 2 \\
4 & 3 & 14 & 12 \\
5 & 2 & 15 & 18 \\
6 & 8 & 16 & 15 \\
7 & 2 & 17 & 2 \\
8 & 7 & 18 & 29 \\
9 & 8 & 19 & 2 \\
10 & 10 & 20 & 26 \\
11 & 2 & 21 & 22 \\
12 & 20 & 22 & 16 \\
\hline
\end{tabular}}
\caption{Number of connected domains of $\NN=1$ vacua for $N_c<23$
  calculated as the number of orbits under the identifications
  (\ref{e.tri})-(\ref{e.trii}).}
\end{table}

So in order to answer the two questions here we have to start from the
set $(N_1,\Dl k, k)$ with $1\leq N_1<N_c$, $0\leq \Dl k< N_c-1$ and $0
\leq k <2N_c-1$ and generate orbits under the transformations
(\ref{e.tri})-(\ref{e.trii}) and their inverses. Each orbit will then
correspond to a connected domain in the space of $\NN=1$ vacua, while
the labels {\em within} a single orbit correspond to different
possible dual descriptions of the same physics. In table 1 we
give the number of orbits (connected domains of $\NN=1$ vacua) for
$N_c \leq 22$. 

\section{Summary and outlook}

The structure of vacua in $\NN=2$ theories broken down to $\NN=1$
appears to be very complex and posesses a rich structure of a number
of connected domains and specific dual descriptions which appear at
the quantum level. In this talk I showed that the use of an explicit exact
solution to the factorization problem may be an efficient tool to
study this structure.

This work may be extended in various directions. One could add
fundamental matter and/or consider different gauge groups (like
orthogonal or symplectic), one might try to understand the
identifications for all {\em three} discrete parameters geometrically
in order to get more insight into the structure of the domains, which
then might be generalized for higher genus. 

\bigskip 

{\bf Acknowledgements: } This work was supported in part by Ministry
of Science and Information Society Technologies 
grants 2P03B08225 (2003-2006), 1P03B02427 (2004-2007) and 1P03B04029
(2005-2008).

\end{document}